# Resistivity reduction of boron-doped multi-walled carbon nanotubes synthesized from a methanol solution containing a boric acid

Satoshi Ishii, Tohru Watanabe, Shinya Ueda, Shunsuke Tsuda, Takahide Yamaguchi and Yoshihiko Takano

National Institute for Materials Science, 1-2-1 Sengen, Tsukuba 305-0047, Japan

Boron-doped multi-walled carbon nanotubes (MWNTs) were synthesized using a methanol solution of boric acid as a source material. Accurate measurements of the electrical resistivity of an individual boron-doped MWNT was performed with a four-point contact, which was fabricated using an electron beam lithography technique. The doped boron provides conduction carriers, which reduces the resistivity of the MWNT.

Carbon nanotubes (CNTs) with low resistivity are especially suited for a variety of applications, such as nano wiring inside large-scale integration (LSI) system and an electrically-conductive coating. The electrical resistivity of the CNTs exhibits metallic or semiconducting properties depending on the chirality of the nanotube [1,2]; therefore controlling the chirarity of single-walled carbon nanotubes (SWNTs) has been attempted by chemical vapor deposition (CVD) using CoMo catalyst to obtain the defined electrical properties [3].

In this study, we introduced conduction carriers into multi-walled carbon nanotubes (MWNTs), which were composed of multiple graphene sheets with different chirality, by boron doping in order to reduce the resistivity. In fact, chemically introduced conduction carriers in carbon allotropes, such as fullerene [4], graphite [5], and diamond [6,7], are found to suppress resistivity, resulting in superconductivity. In particular, heavily boron doped diamond is found to exhibit superconductivity, even though it is a band insulator with a wide bandgap [8]. It was thus speculated that boron is an appropriate element that can be used to introduce conduction carriers effectively into MWNTs. In addition, MWNTs are likely to exhibit superconductivity [9], although a one-dimensional material should not exhibit superconductivity from the quantum transport point of view [10,11]. MWNTs have multiple-dimensionality due to the interlayer interaction [12]. If a superconducting MWNT could be wired inside an LSI using a selective growth technique, a high density and low-power-consumption LSI could be produced.

In order to synthesize the boron-doped MWNTs, we have developed a synthesis method which uses a methanol solution of boric acid as a source material. Furthermore, a typical thermal CVD system, which was simply composed of an electric furnace internally equipped with a quartz tube, a scroll pump, and a flask filled with the source solution [13], was employed to produce the nanotubes. Thus, we can grow boron-doped MWNTs with greater ease, safety, and inexpensively. Boron-doped MWNTs were grown on an Si substrate. Prior to growth, the surface of the Si substrate was pre-coated with iron acetate as a metal catalyst to activate MWNT growth [13]. The pre-coated Si substrate was then placed inside the quartz tube, and heated to 950 $^o$C at a rate of 46 $^o$C/min under a pressure of 10 Torr. The substrate was kept at 950 $^o$C for 15 min to obtain pure iron from iron oxide, and subsequently the temperature was decreased to 730 $^o$C at a rate of 14.7 $^o$C/min. At 730 $^o$C, the vaporized source solution was introduced into the quartz tube at a constant pressure of 200 Torr to grow boron-doped MWNTs. After four hours, the introduction was stopped, and the substrate was cooled down to room temperature. The boron-doped MWNTs were synthesized from two source solutions containing 1.0 atm% and 2.0 atm% of boron. In this paper, boron-doped MWNTs grown at these concentrations are denoted as "1.0 atm%" and "2.0 atm%", respectively. However, the boron concentration of the MWNTs is different from that of the source solutions.

As-grown boron-doped MWNTs on the surface of the Si substrate were observed using a field emission scanning electron microscope (FE-SEM) at an accelerating voltage of 5.0 kV. The detailed structure of the boron-doped MWNT, on a carbon-coated micro-grid, was observed using a transmission electron microscope (TEM) at an accelerating voltage of 400 kV. In order to evaluate the effect of the boron-doping on the structural properties of the MWNTs, Raman spectra of both non-doped (commercially available: denoted as "0 atm%") and the doped samples were obtained using a micro-Raman system with an Ar-ion laser excitation of 514.5 nm in atmosphere, and at room temperature.

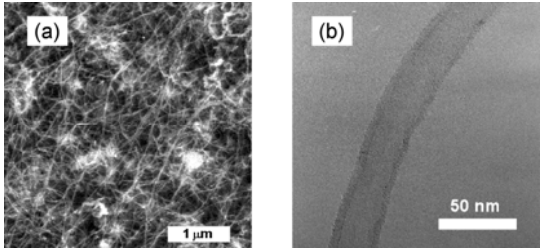

Fig. 1. (a) SEM image of as-grown boron-doped MWNTs (2.0 atm %) on a surface of an Si substrate. (b) TEM image of a boron-doped MWNT (2.0 atm%) on a micro-grid. The image represents the detailed structure of the MWNT.

The temperature dependence of resistivity in each individual MWNT was measured using the four-point method, which provides accurate resistivity measurements. Four small-sized electrodes of several hundred nanometers in width were fabricated on a target MWNT using electron beam (EB) lithography. The MWNT and electrodes were placed on an $SiO_2$ insulating layer (500 nm thick) on an Si wafer. Ti/Au (50/100 nm thick) was deposited as metal electrodes on the MWNT. The measurements were carried out in a liquid helium cooling system from room temperature down to 2 K.

Figure 1(a) is a typical SEM image of the as-grown boron-doped MWNTs (2.0 atm%) on an Si substrate. 1.0 atm% samples were found to be similar to the 2.0 atm% samples. Using the current technique, a large number of boron-doped MWNTs were produced in a single run. Figure 1(b) is a TEM image of the detailed structure of the boron-doped MWNT (2.0 atm%). The boron-doped MWNT was found to be composed of multiple graphene sheets with the inner and outer diameters of 20 nm and 35 nm, respect-ively. The number of graphene sheet layers is thus estimated to be 22, by taking the typical interlayer distance as 3.35 Å.

We obtained Raman spectra of three kinds of MWNTs with different boron doping levels. Typical results are shown in Fig. 2 (a). We found two distinct peaks around 1600 $cm^{-1}$ and 1350 $cm^{-1}$ for all the MWNTs; labeled as G-band and D-band, respectively. The G-band is due to lattice vibrations of the C-C bond, and the D-band is due to the defects in the MWNTs [14]. The peak positions of the G-band are plotted as a function of the boron concentration of the source solution in Fig. 2(b). The peak positions shift to the higher Raman side linearly for higher boron concentrations. The upshift indicates that hole carriers have been transferred from boron to the MWNTs. The charge transfer shortens the C-C bond increasing the force constant, and thus enhances the lattice frequency of the MWNTs [15]. The full-width at half-maximum (FWHM) of the G-band is plotted as a function of boron-concentration of the source solution in Fig. 2(c). The FWHM of the G-band increases as the boron concentration increases. The broadening of the peaks may be attributed to the inhomogeneity of the boron concentration in the MWNTs [14]. Although we have no information about the concentrations of doped boron atoms and defects, and the distribution of chirality, it is clear that the MWNTs contain boron, and the amount of the boron increases when the boron concentration of the source solution is increased.

Figure 3 is a plot of the normalized resistivity, $\rho/\rho_{RT}$, as a function of temperature, where $\rho_{RT}$ is the resistivity at room temperature. Clearly, the general trend indicates that the temperature dependence of $\rho/\rho_{RT}$ of the boron-doped MWNTs is much weaker than that of a typical semiconductor with a thermally activated conduction process. Such weak temperature dependence of resistivity was also reported by Zhou et al. [16]. In their experiments, linear temperature dependence of conductance was observed in the MWNTs. They explain the transport process by the model, where the total number of conduction channels is increased by increasing temperature and the number of conduction carriers. Our experimental result is also similar to the Zhou's, and therefore that may be explained by their model. Our result that $\rho/\rho_{RT}$ is well suppressed for increasing concentrations of the boron dopant is consistent with their model. In other words, the total number of conduction channels at 0 K is increased when the number of conduction carriers is increased by the boron doping [16].

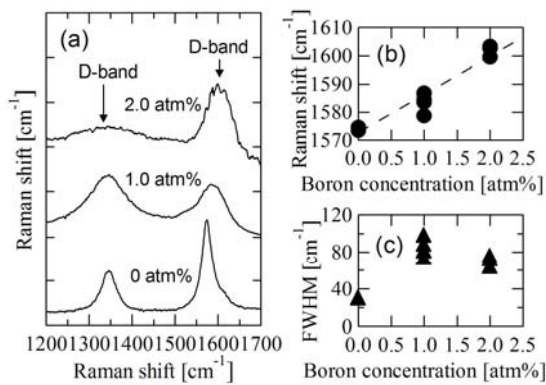

Fig. 2. (a) Raman spectra of boron-doped MWNTs; 0 atm%, 1.0 atm%, and 2.0 atm%. The spectra were obtained by using 514.5 nm laser excitation. (b) Boron concentration dependence of peak positions of G-band. (c) Boron concentration dependence of FWHM of the G-band.

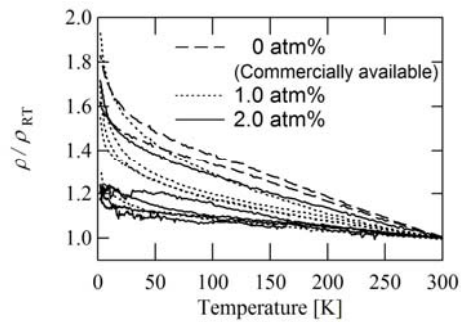

Fig. 3. Temperature dependence of normalized resistivity $\rho/\rho_{RT}$ in the MWNTs. The resistivity $\rho$ was normalized by the room temperature resistivity $\rho_{RT}$. The dashed curves are for 0 atm%, the dotted lines, 1.0 atm%, and the solid, 2.0 atm%.

Figure 4 is a plot of the resistivity of 2.0 atm% samples as a function of temperature. The increases of resistivity are small, as the resistivity at room temperature becomes lower. The inset of Fig. 4 is a magnification of the low temperature region of a single sample (circled), which shows the lowest resistivity at room temperature. In particular, a sharp resistivity drop is observed below 12 K. The boron should affect the electronic conduction mechanism of the MWNT, since such a sharp drop was observed only in the boron-doped MWNT. Transport measurements on the heavily boron-doped MWNTs in various magnetic fields at low temperatures using a 3He cryostat system would provide useful information.

In summary, we have reduced the resistivity of MWNTs by boron doping. Boron-doped MWNTs were produced from a methanol solution of boric acid. The resistivity of the boron-doped MWNTs was found to decrease for higher boron concentrations of the source solution. Doped boron produces conduction carriers effectively in the MWNTs. We obtained one sample which exhibited a resistivity drop below 12 K. This unconventional conduction was found only in the boron-doped MWNT. We will attempt to produce samples with greater concentrations of boron to reduce the resistivity further and investigate superconductivity in the MWNT.

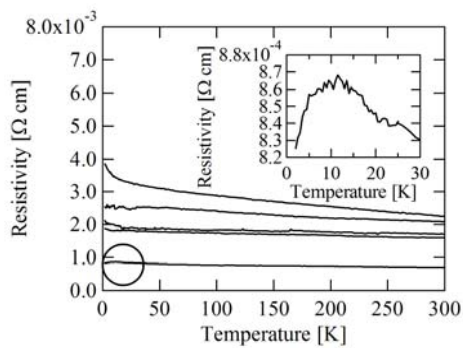

Fig. 4. Temperature dependence of resistivity only for 2.0 atm% samples. Circled sample which shows the lowest resistivity at room temperature is in the inset: the temperature dependence of resistivity below 30 K.